# No limit on the energy extracted from a Dirac quantum state

by

Dan Solomon

Rauland-Borg Corporation
3450 W. Oakton Street
Skokie, IL 60076

Phone: 847-324-8337
Email: **dan.solomon@rauland.com**

PACS 11.10.-z

(Oct. 12, 2002)

**Abstract**

When a Dirac quantum state interacts with an applied electric field the energy of the quantum state changes. It is generally assumed that there is a maximum limit on the amount of energy that can be extracted from a Dirac quantum state, due to its interaction with an electric field. In this article it is shown that this assumption is not correct and that, for a properly applied electric field, an arbitrarily large amount of energy can be extracted from the quantum state.

# I. Introduction

In Dirac field theory the free field energy is the energy of the quantum state when the interactions are turned off. It is generally assumed that there is a lower bound to the free field energy, which is the energy of the vacuum state. However it has been recently shown [1] that for Dirac field theory to be gauge invariant there must be no lower bound to the free field energy.

The investigation undertaken in [1] was motivated by the problem of gauge invariance. Even though Dirac field theory is assumed to be gauge invariant it is well known that when the polarization tensor is calculated, using perturbation theory, the result is not gauge invariant (see Chapt. 14 of [2], Sect. 22 of [3], Chapt. 5 of [4], and [5]). The non-gauge invariant terms must be isolated and removed from the results of the calculation in order to obtain the correct gauge invariant result. This may involve some form of regularization, where other functions that are introduced have the correct behavior so that the non-gauge invariant terms are cancelled. However there is no physical explanation for introducing these functions. They are merely mathematical devices used to force the desired (gauge invariant) result [5].

It was shown in [1] that it is possible to modify the standard definition of the vacuum state so that there is no lower bound to the free field energy. When this is done perturbation theory will be gauge invariant. If there is no lower bound to the free field energy this suggests that it is worth examining the following problem: Given some initial Dirac quantum state, is there any limit on the amount of energy that can be extracted from the quantum state due to its interaction with an applied electric field? It will be shown that there is no such limit for a quantum state that meets certain initial conditions.

In the following discussion we will work in the Heisenberg picture and assume that $\hbar = c = 1$. In the Heisenberg picture the state vector $|\Omega\rangle$ is constant in time and the field operator $\hat{\psi}(\vec{x},t)$ is time dependent and obeys the following equation [2][6],

$$i\frac{\partial \hat{\psi}(\vec{x},t)}{\partial t} = H_D \hat{\psi}(\vec{x},t) \tag{1}$$

where,

$$H_D = H_0 - q\vec{\alpha} \cdot \vec{A} + qA_0 \tag{2}$$

and

$$H_0 = -i\vec{\alpha} \cdot \vec{\nabla} + \beta m \tag{3}$$

In the above expression q and m are the charge and mass of the electron, respectively, and the 4x4 matrices $\vec{\alpha}$ and $\beta$ are defined in [2]. The pair $(A_0, \vec{A})$ is the electric potential and is taken to be an unquantized, real valued quantity. As shown in the above expressions the coupling between the electromagnetic field and the Dirac quantum state is through the electric potential.

The free field energy is the energy of the state vector $|\Omega\rangle$ when the electric potential is zero. It is defined by,

$$\xi_f(t) = \langle \Omega | \hat{H}_0(t) | \Omega \rangle - \varepsilon_r \tag{4}$$

where,

$$\hat{H}_0(t) = \int \left( \hat{\psi}^\dagger(\vec{x},t) H_0 \hat{\psi}(\vec{x},t) \right) d\vec{x} \tag{5}$$

and where $\varepsilon_r$ is a renormalization constant which is chosen so that the free field energy of the vacuum state is zero.

## II. Extracting Energy

If the electric potential is zero then $\xi_f(t)$ is constant in time. If a non-zero electric potential is applied for a period of time then, in general, the value of $\xi_f(t)$ will change. After the electric potential is removed the change in the energy of the quantum state is the difference between the new value of $\xi_f$ and the original value.

Now consider the change in energy of the quantum state when it is acted on by the following electric potential,

$$\left(A_0(\vec{x},t), \vec{A}(\vec{x},t)\right) = 0 \text{ for } t < 0$$

$$\left(A_0(\vec{x},t), \vec{A}(\vec{x},t)\right) = \left(\frac{\partial \chi(\vec{x},t)}{\partial t}, -\vec{\nabla}\chi(\vec{x},t)\right) \text{ for } 0 \le t < t_a \qquad (6)$$

$$\left(A_0(\vec{x},t), \vec{A}(\vec{x},t)\right) = 0 \text{ for } t \ge t_a$$

where $\chi(\vec{x},t)$ is an arbitrary real valued function subject to the following condition at $t=0$,

$$\chi(\vec{x},0) = 0; \quad \left.\frac{\partial \chi(\vec{x},t)}{\partial t}\right|_{t=0} = 0 \qquad (7)$$

From the above expressions the electric potential is zero at time $t=0$ and time $t_b > t_a$. Therefore the change in energy from t=0 to $t_b > t_a$ is given by the difference in the free field energies, i.e.,

$$\Delta \xi = \xi_f(t_b) - \xi_f(0) \qquad (8)$$

Use (6) in (1) along with (2) to show that the field operator satisfies the following equations,

$$i\frac{\partial \hat{\psi}(\vec{x},t)}{\partial t} = \left(H_0 + q\vec{\alpha}\cdot\vec{\nabla}\chi(\vec{x},t) + q\frac{\partial \chi(\vec{x},t)}{\partial t}\right)\hat{\psi}(\vec{x},t) \text{ for } 0 \leq t < t_a \qquad (9)$$

and,

$$i\frac{\partial \hat{\psi}(\vec{x},t)}{\partial t} = H_0 \hat{\psi}(\vec{x},t) \text{ for } t \geq t_a \qquad (10)$$

The solution to (9) is,

$$\hat{\psi}(\vec{x},t) = e^{-iq\chi(\vec{x},t)} e^{-iH_0 t} \hat{\psi}(\vec{x},0) \text{ for } 0 \leq t < t_a \qquad (11)$$

This can be seen from substituting (11) into (9) and using the fact that,

$$H_0\left(e^{-iq\chi(\vec{x},t)} e^{-iH_0 t} \hat{\psi}(\vec{x},0)\right) = \left(-i\vec{\alpha}\cdot\vec{\nabla} + \beta m\right)\left(e^{-iq\chi(\vec{x},t)} e^{-iH_0 t} \hat{\psi}(\vec{x},0)\right)$$
$$= e^{-iq\chi(\vec{x},t)}\left(-q\vec{\alpha}\cdot\vec{\nabla}\chi(\vec{x},t) + H_0\right) e^{-iH_0 t} \hat{\psi}(\vec{x},0) \qquad (12)$$

and,

$$i\frac{\partial}{\partial t}\left(e^{-iq\chi(\vec{x},t)} e^{-iH_0 t} \hat{\psi}(\vec{x},0)\right) = e^{-iq\chi(\vec{x},t)}\left(q\frac{\partial \chi(\vec{x},t)}{\partial t} + H_0\right) e^{-iH_0 t} \hat{\psi}(\vec{x},0) \qquad (13)$$

The solution to (10) is,

$$\hat{\psi}(\vec{x},t) = e^{-iH_0(t-t_a)} \hat{\psi}(\vec{x},t_a) \text{ for } t \geq t_a \qquad (14)$$

Since (1) is a first order differential equation the condition at $t = t_a$ that must be satisfied is,

$$\hat{\psi}(\vec{x},t_a + \delta)\underset{\delta \to 0}{=} \hat{\psi}(\vec{x},t_a - \delta) \qquad (15)$$

Using (11) and (14) this yields,

$$\hat{\psi}(\vec{x},t_a) = e^{-iq\chi(\vec{x},t_a)} e^{-iH_0 t_a} \hat{\psi}(\vec{x},0) \qquad (16)$$

Use this in (14) to obtain,

$$\hat{\psi}(\vec{x},t) = e^{-iH_0(t-t_a)} e^{-iq\chi(\vec{x},t_a)} e^{-iH_0 t_a} \hat{\psi}(\vec{x},0) \text{ for } t > t_a \qquad (17)$$

To evaluate the energy at $t_b > t_a$ use this result in (5) to obtain,

$$\hat{H}_0(t_b)\Big|_{t_b>t_a} = \int \left( \hat{\psi}^\dagger(\vec{x},0) e^{iH_0 t_a} e^{iq\chi(\vec{x},t_a)} e^{iH_0(t_b-t_a)} H_0 e^{-iH_0(t_b-t_a)} e^{-iq\chi(\vec{x},t_a)} e^{-iH_0 t_a} \hat{\psi}(\vec{x},0) \right) d\vec{x} \tag{18}$$

This readily yields,

$$\hat{H}_0(t_b)\Big|_{t_b>t_a} = \int \left( \hat{\psi}^\dagger(\vec{x},0) e^{iH_0 t_a} e^{iq\chi(\vec{x},t_a)} H_0 e^{-iq\chi(\vec{x},t_a)} e^{-iH_0 t_a} \hat{\psi}(\vec{x},0) \right) d\vec{x} \tag{19}$$

Use (12) in (19) to obtain,

$$\hat{H}_0(t_b)\Big|_{t_b>t_a} = \int \left( \hat{\psi}^\dagger(\vec{x},0) e^{iH_0 t_a} \left( -q\vec{\alpha}\cdot\vec{\nabla}\chi(\vec{x},t_a) + H_0 \right) e^{-iH_0 t_a} \hat{\psi}(\vec{x},0) \right) d\vec{x} \tag{20}$$

This yields,

$$\hat{H}_0(t_b)\Big|_{t_b>t_a} = \int \begin{Bmatrix} \left( \hat{\psi}^\dagger(\vec{x},0) H_0 \hat{\psi}(\vec{x},0) \right) \\ -q\left( \hat{\psi}_0^\dagger(\vec{x},t_a) \vec{\alpha}\hat{\psi}_0(\vec{x},t_a) \right)\cdot\vec{\nabla}\chi(\vec{x},t_a) \end{Bmatrix} d\vec{x} \tag{21}$$

where,

$$\hat{\psi}_0(\vec{x},t_a) = e^{-iH_0 t_a}\hat{\psi}(\vec{x},0) \tag{22}$$

Use (5) in (21) to obtain,

$$\hat{H}_0(t_b)\Big|_{t_b>t_a} = \hat{H}_0(0) - \int \hat{\vec{J}}(\hat{\psi}_0(\vec{x},t_a))\cdot\vec{\nabla}\chi(\vec{x},t_a) d\vec{x} \tag{23}$$

where,

$$\hat{\vec{J}}(\hat{\psi}) \equiv q\hat{\psi}^\dagger \vec{\alpha}\hat{\psi} \tag{24}$$

The quantity $\hat{\vec{J}}(\hat{\psi})$ is the current operator. Use (23) to obtain,

$$\langle\Omega|\hat{H}_0(t_b)|\Omega\rangle\Big|_{t_b>t_a} = \langle\Omega|\hat{H}_0(0)|\Omega\rangle - \int \langle\Omega|\hat{\vec{J}}(\hat{\psi}_0(\vec{x},t_a))|\Omega\rangle\cdot\vec{\nabla}\chi(\vec{x},t_a) d\vec{x} \tag{25}$$

Rearrange terms and use (4) and (8) to obtain,

$$\Delta\xi = -\int \langle\Omega|\hat{\vec{J}}(\hat{\psi}_0(\vec{x},t_a))|\Omega\rangle \cdot \vec{\nabla}\chi(\vec{x},t_a)d\vec{x} \tag{26}$$

Integrate by parts and assume reasonable boundary conditions to obtain,

$$\Delta\xi = \int \chi(\vec{x},t_a)\vec{\nabla}\cdot\langle\Omega|\hat{\vec{J}}(\hat{\psi}_0(\vec{x},t_a))|\Omega\rangle d\vec{x} \tag{27}$$

The quantity $\langle\Omega|\hat{\vec{J}}(\hat{\psi}_0(\vec{x},t_a))|\Omega\rangle$, in the above expression, is independent of $\chi(\vec{x},t_a)$. Therefore we can change the value of $\chi(\vec{x},t_a)$ without affecting the value of $\langle\Omega|\hat{\vec{J}}(\hat{\psi}_0(\vec{x},t_a))|\Omega\rangle$. Assume that the divergence of $\langle\Omega|\hat{\vec{J}}(\hat{\psi}_0(\vec{x},t_a))|\Omega\rangle$ is non-zero, i.e.,

$$\vec{\nabla}\cdot\langle\Omega|\hat{\vec{J}}(\hat{\psi}_0(\vec{x},t_a))|\Omega\rangle \neq 0 \text{ for some } \vec{x} \tag{28}$$

Then we can always find a $\chi(\vec{x},t_a)$ that makes $\Delta\xi$ a negative number with an arbitrarily large magnitude. For example, let,

$$\chi(\vec{x},t_a) = -f\vec{\nabla}\cdot\langle\Omega|\hat{\vec{J}}(\hat{\psi}_0(\vec{x},t_a))|\Omega\rangle \tag{29}$$

where f is a constant. Use this in (27) to obtain,

$$\Delta\xi = -f\int \left(\vec{\nabla}\cdot\langle\Omega|\hat{\vec{J}}(\hat{\psi}_0(\vec{x},t_a))|\Omega\rangle\right)^2 d\vec{x} \tag{30}$$

The integral, in the above expression, is always positive. Therefore $\Delta\xi \to -\infty$ as $f \to +\infty$. This means that there is no limit on the amount of energy the can be extracted from the Dirac quantum state provided that we start with a state vector that satisfies equation (28).

### **III. Conclusion**

There is, in principle, no limit on the amount of energy that can be extracted from a Dirac state due to its interaction with the electric potential given above. This result also shows that there is no lower bound to the free field energy. This is because, from (8), the final free field energy $\xi_f(t_b)$ is less than the initial free field energy $\xi_f(0)$ by an arbitrarily large amount as $\Delta\xi \to -\infty$. This confirms the conclusion of previous work [1] where it was shown, by different means, that there can be no lower bound to the free field energy in Dirac field theory.